\documentclass[fleqn,10pt]{wlscirep}
\usepackage[utf8]{inputenc}
\usepackage[T1]{fontenc}
\DeclareUnicodeCharacter{03B3}{$\gamma$}
\DeclareUnicodeCharacter{02BC}{*}
\definecolor{dkblue}{RGB}{54, 86, 169}

\newcommand{\araa}{Annu. Rev. Astron. Astrophys.}   
\newcommand{\aj}{Astron. J.}   
\newcommand{\apj}{Astrophys. J.}   
\newcommand{\apjl}{Astrophys. J. Lett.}   
\newcommand{\apjs}{Astrophys. J. Suppl. Ser.}   
\newcommand{\ao}{Appl. Opt.}   
\newcommand{\aap}{Astron. Astrophys.}   
\newcommand{\mnras}{Mon. Not. R. Astron. Soc.}   
\newcommand{\nat}{Nature} 
\newcommand{\prd}{Phys. Rev. D}   
\newcommand{\pasp}{Publ. Astron. Soc. Pac.}   

\title{Prompt-to-afterglow transition of optical emission in a long gamma-ray burst consistent with a fireball}
\author[1]{Liping Xin}
\author[1]{Xuhui Han}
\author[1]{Huali Li}
\author[2,3] {Bing Zhang}
\author[4]{Jing Wang}
\author[5]{Damien Turpin}
\author[4]{Xing Yang}
\author[1]{Yulei Qiu}
\author[4]{Enwei Liang}
\author[7,8]{Zigao Dai}
\author[1]{Hongbo Cai}
\author[1]{Xiaomeng Lu}
\author[8,9]{Xiang-Yu Wang}
\author[1]{Lei Huang}
\author[4]{Xianggao Wang}
\author[1]{Chao Wu}
\author[11,12]{He Gao}
\author[8,9]{Jia Ren}
\author[4]{Lulu Zhang}
\author[10]{Yuangui Yang}
\author[1,6]{Jingsong Deng}
\author[1,6]{Jianyan Wei}

\affil[1]{CAS Key Laboratory of Space Astronomy and Technology, National Astronomical Observatories, Chinese Academy of Sciences, Beijing 100101, China.}
\affil[2] {Nevada Center for Astrophysics, University of Nevada, Las Vegas, NV 89154, USA}
\affil[3]{Department of Physics and Astronomy, University of Nevada, Las Vegas, NV 89154, USA}
\affil[4]{Guangxi Key Laboratory for Relativistic Astrophysics, School of Physical Science and Technology, Guangxi University, Nanning 530004, China}
\affil[5]{Universit$\acute{e}$ Paris-Saclay, Universit$\acute{e}$ Paris Cit$\acute{e}$, CEA, CNRS, AIM, 91191, Gif-sur-Yvette, France}
\affil[6]{School of Astronomy and Space Science, University of Chinese Academy of Sciences, Beijing, Peopleʼs Republic of China}
\affil[7]{Department of Astronomy, University of Science and Technology of China, Hefei 230026, Peopleʼs Republic of China}
\affil[8]{School of Astronomy and Space Science, Nanjing University, Nanjing 210093, China}
\affil[9]{Key Laboratory of Modern Astronomy and Astrophysics (Nanjing University), Ministry of Education, Nanjing 210093, Peopleʼs Republic of China}
\affil[10]{School of Physics and Electronic Information, Huaibei Normal University, Huaibei 235000, Anhui Province, China}
\affil[11]{Institute for Frontier in Astronomy and Astrophysics, Beijing Normal University, Beijing 102206, China}
\affil[12]{Department of Astronomy, Beijing Normal University, Beijing 100875, People’s Republic of China}

\begin{abstract}
{\bf 
Long gamma-ray bursts (GRBs), which signify the end-life collapsing of very massive stars, are produced by extremely relativistic jets colliding into circumstellar medium. Huge energy is released both in the first few seconds, namely the internal dissipation phase that powers prompt emissions,  and in the subsequent self-similar jet-deceleration phase that produces afterglows observed in broad-band electromagnetic spectrum. However, prompt optical emissions of GRBs have been rarely detected, seriously limiting our understanding of the transition between the two phases. Here we report detection of prompt optical emissions from a gamma-ray burst (i.e. GRB 201223A) using a dedicated telescope array with a high temporal resolution and a wide time coverage. The early phase coincident with prompt $\gamma$-ray emissions show a luminosity in great excess with respect to the extrapolation of $\gamma$-rays, while the later luminosity bump is consistent with onset of the afterglow. The clearly detected transition allows us to differentiate physical processes contributing to early optical emissions and to diagnose the composition of the jet.
}

\end{abstract}
\begin{document}

\flushbottom
\maketitle
%
%
\section*{Main}
A long-duration gamma-ray bursts (GRB) is produced by the collapse of a massive star into a black hole or a rapidly spinning, highly magnetized neutron star, with enormous energy emitted in the entire electromagnetic spectrum in the first few seconds from a jet moving towards Earth with a speed close to the speed of light\cite{2018pgrb.book.....Z}. Because of the  relativistic effects, the jet transitions from the internal dissipation phase that produces prompt emission to the self-similar deceleration phase that powers broad-band afterglow within tens to hundreds of seconds in the observer frame. Even though such a transition has been widely observed in the X-ray band\cite{2006ApJ...642..354Z,2006ApJ...642..389N}, it is not well studied in the optical band because of the sparse coverage of optical emission throughout the entire gamma-ray emission phase using telescopes with both large field of view and high sensitivity. Bright optical flashes associated with prompt emission have been observed in several GRBs\cite{2005Natur.435..178V,2006Natur.442..172V,2008Natur.455..183R}, but their immediate transition to early afterglow was only studied in limited cases\cite{2006Natur.442..172V}. Some other early optical observations often show a significant emission component from a bright reverse shock\cite{1999Natur.398..400A,2014Sci...343...38V,2017Natur.547..425T}, which smears the clear transition signature from prompt to afterglow emission. 
The Ground-based Wide-angle Camera array (GWAC)\cite{2016arXiv161006892W} is a wide-angle (a field of view of 2200 square degrees built currently),
moderate sensitivity (detection limit around 16 magnitudes) optical facility that can monitor GRB prompt emission phase with a high temporal cadence (integration times of 10 seconds), 
aiming to capture the emergence of optical flares accompanying the prompt high energy emission, as well as offer a glimpse of the clear transition from the prompt to afterglow phase. 

On 23 December 2020, 
GRB 201223A triggered the Bursts Alert Telescope (BAT) onboard the {\em Neil Gehrels Swift Observatory}\cite{2004ApJ...611.1005G} and the Gamma-Ray Burst Monitor (GBM) of the {\em Fermi Gamma-Ray Telescope}\cite{2009ApJ...702..791M} at 17:58:26 Universal Time (UT)\cite{2020GCN.29158....1G,2020GCN.29161....1W}. 
The duration of gamma-ray emission  was $\sim$29 sec with a 15-350 keV fluence of approximately 1.64$\times10^{-6}$  erg/cm$^2$ based on a Cutoff-PL spectral model\cite{2020GCN.29158....1G}. 
The afterglow was detected by Swift X-ray telescope (XRT) and UV/optical telescope (UVOT)\cite{2020GCN.29158....1G},  with a bright optical counterpart of $16.76\pm0.06$ magnitude  identified in an UVOT-$u$ filter image at about 292 seconds after the trigger. No spectroscopic redshift was reported.
If we conservatively place the $Ly {\alpha}$ limit at the middle of the UVOT-$u$ bandpass (346.5 nm\cite{2008MNRAS.383..627P}), the redshift of GRB 201223A would be smaller than 1.85.

\begin{figure}
\centering
\includegraphics[width=1\linewidth]{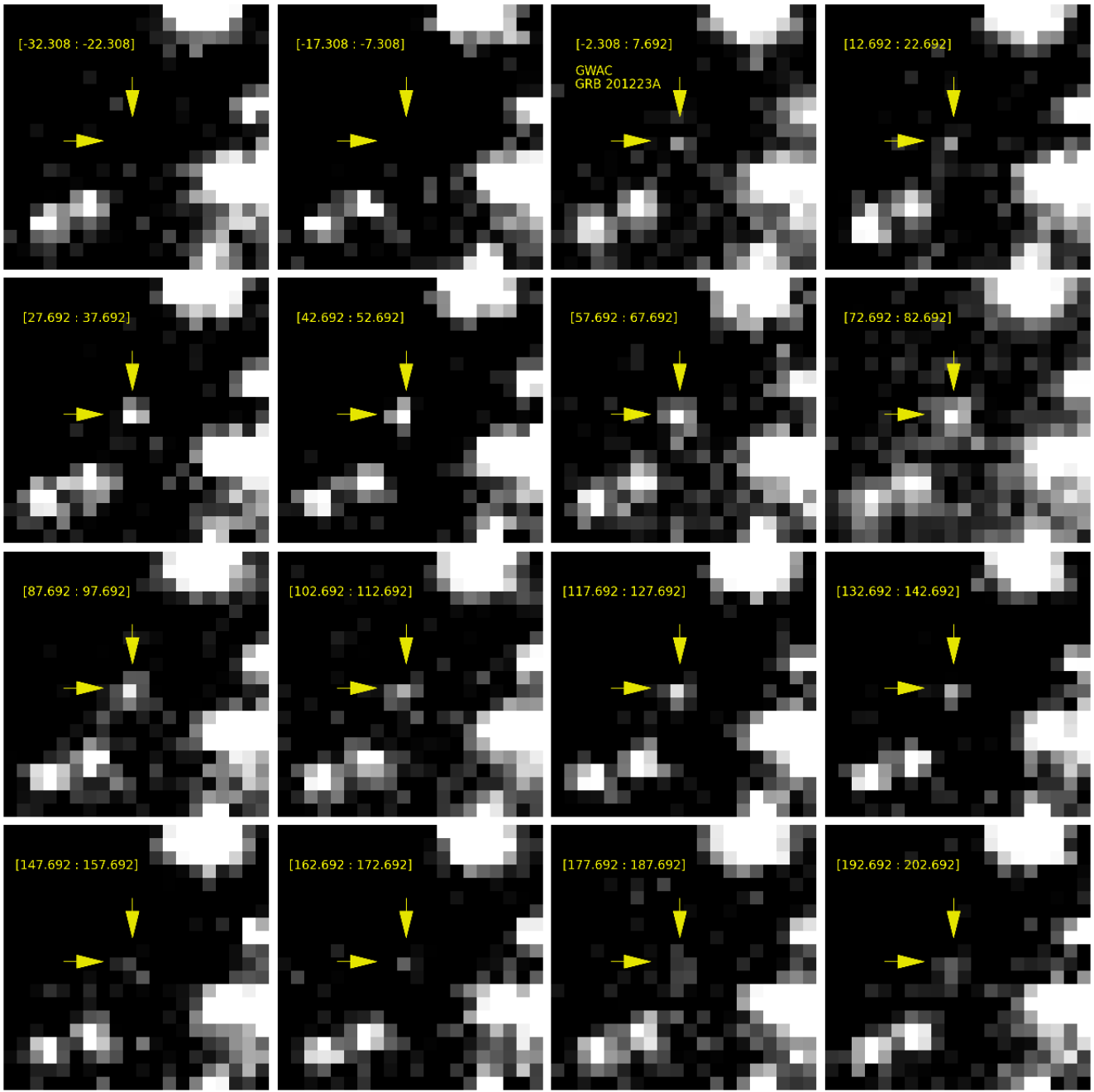}
\caption{\textbf{GRB 201223A was observed by GWAC before, during, and after the GRB providing the transition from prompt to afterglow emission and insight into the composition of GRB jets.} The images show the GRB location in sixteen consecutive frames obtained by the 0.18-m ground based wide angle cameras, owned by the National Astronomical Observatories, CAS, China, and located at the Xinglong observatory. Each GWAC telescope uses an unfiltered 4000 * 4000 pixel back-illuminated charge-coupled device (CCD) camera and typical achieves a 3$\sigma$ limiting magnitude of R~15.4 mag for 10 second exposure.The displayed images span the time interval from 17:57:53 to 18:01:23 UT on 23 December 2021. The arrows points  the location of the GRB optical counterpart (right ascension 08h 51 min 09.51s, declination $+71^\circ 10^\prime 47.4^{\prime \prime}$(J2000)). The right and upper directions denote East and North, respectively.  The effective exposure time range is labeled in each frame, which is relative to the Swift/BAT trigger time in seconds. }
\label{fig:findingchart}
\end{figure}

A large patch of the sky area covering GRB 201223A was monitored continuously by GWAC with a cadence of 15 seconds, lasting for 
95 minutes from 16:48:31.9 to 18:23:41.9 UT covering 70 minutes prior to the GRB trigger through 25 minutes post-trigger at the Xinglong observatory in China. These intense observations covered the entire process of the event. 
The emergence of a fast transient within its field of view at the location of GRB 201223A was serendipitously detected by one of the GWAC instruments, which is temporally coincident with the prompt gamma-ray pulses of the GRB (see Methods).
The optical counterpart was detected in 16 consecutive images as shown in Fig.\ref{fig:findingchart}, which covered the total duration of high energy emission, providing a rare glimpse of the temporal activity in the visual wavelengths around the onset of the gamma-ray burst. 

The event was also responded automatically by the F60A telescope after the alert message was received, resulting in optical observations starting at 45 seconds  post-$\gamma$-ray-trigger and stopping at the time of $\sim$3 ks when the event faded down to the limiting magnitude of $\sim$18. 

\begin{figure}
\centering
\includegraphics[width=1.0\linewidth]{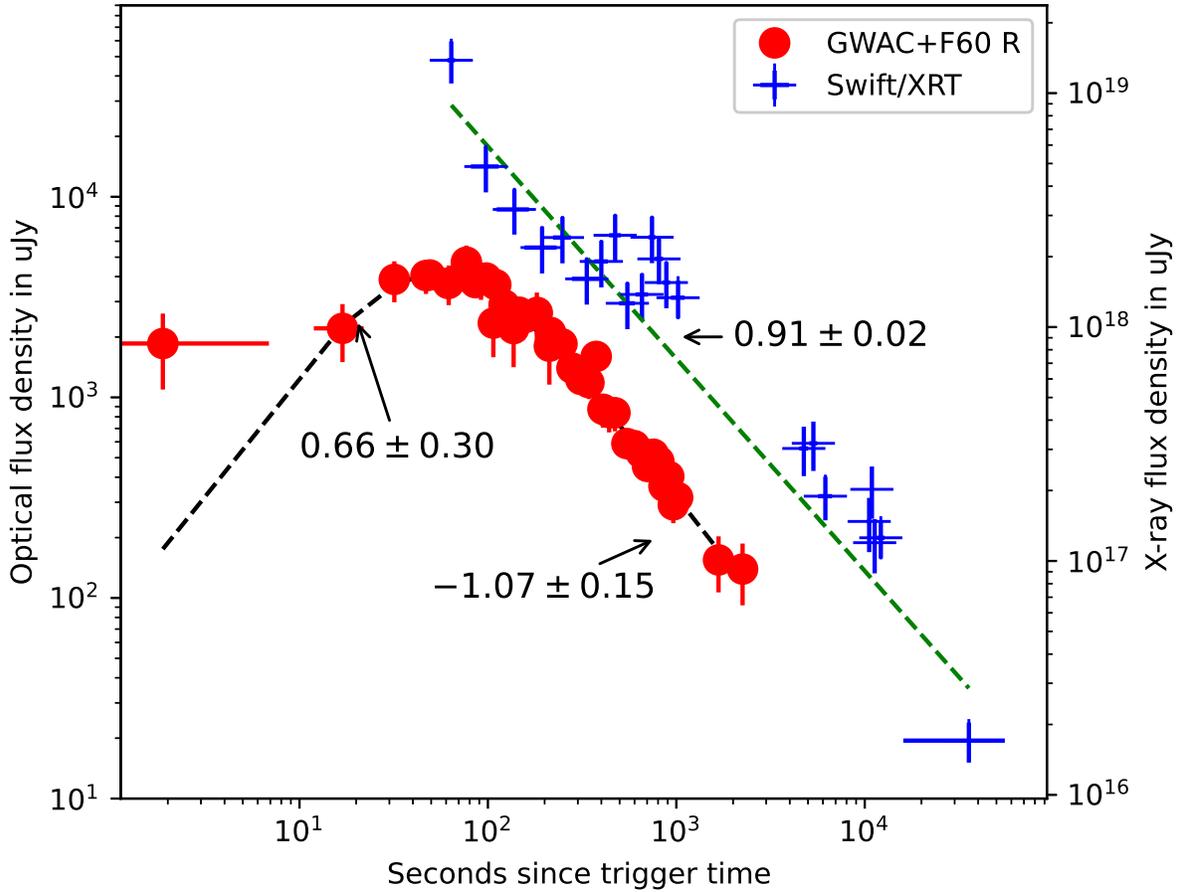}
\caption{ \textbf{Optical and X-ray light curves of GRB 201223A and their modeling}. Optical data are derived from GWAC and GWAC-F60A shown in red. The X-ray flux in blue are measured by XRT instrument onboard Swift satellite. The x-axis is the time after the trigger in seconds. 
The black dashed line shows the well fitting during the temporal range [16s:3000s] with a broken power-law model.  
All the error bars denote the 1$\sigma$ statistical errors. 
From the second GWAC measurement,the flux is brightening with a slope of $\alpha_1 =0.66\pm0.30$ before the peak time of  $t_m = 52\pm30$sec. The optical emission enters a decay phase as a single power law with an index of $\alpha_2 =-1.07\pm0.15$. The reduced $\chi^{2} =0.67$ with a degree of freedom (DoF) of 28. The green dash line reviews the best fitting for the X-ray light curve with a slope of 0.91$\pm$0.02 with an $\chi^{2}$/DoF = 77.0/20.
}
\label{fig:optical_modeling}
\end{figure}

The optical, X-rays and gamma-ray light curves are displayed in Fig.\ref{fig:lc} .
We obtain the BAT and the XRT light curves from the XRT
light curve and spectral repository\cite{2009MNRAS.397.1177E}.
The observed 15-150 keV flux measured by Swift BAT is integrated with a  0.256 s temporal revolution, showing several spikes during its activities.
The main emission after the trigger time lasts for about 10 seconds and some weak precursors are also observed.
The X-ray lightcurve shows a single power law decay with an index of $\alpha_x \sim 0.91\pm0.02$(Fig.\ref{fig:optical_modeling}). During the decay phase, there is a weak signature of rebrightening which indicates a re-activation of the central engine. During and before the faint gamma-ray precursor, no optical emission was detected in our images down to a limit of $\sim$15.3 mag in $R$ band. 
After its sudden appearance during the main gamma-ray prompt emission phase, the optical brightness remains briefly constant during the first two frames and then rises rapidly to the peak of 14.5 mag in the R-band with a power law index of $0.66\pm0.30$. It reaches the peak at $52\pm 34$ s and then transitions to a decay phase with a decay index of $\sim -1.07\pm0.15$(Fig.\ref{fig:optical_modeling}), which extends all the way down to the sensitivity limit of the F60A at $\sim$ 3000 s.

\begin{figure}
\centering
\includegraphics[width=1.0\linewidth]{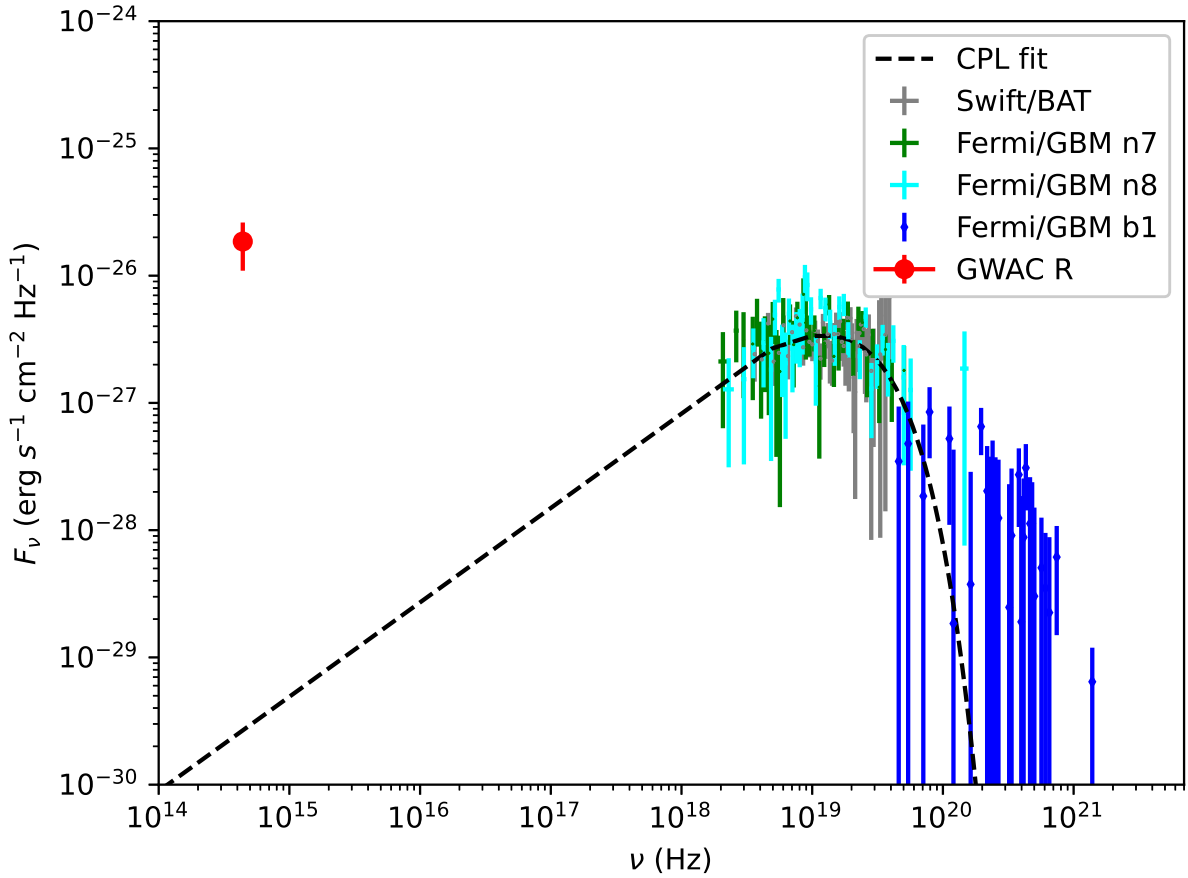}
\caption{
\textbf{Broadband spectra of the prompt phase in GRB 201223A.}
The red solid data was observed by GWAC. The gray data were measured by Swift/BAT and other data were derived from  Fermi/GBM. The time duration for all the data is from -2.31 seconds to 7.69 seconds post the burst. The black dash line denotes the best fitting model after the joint analysis of the GWAC data, Swift/BAT data and  Fermi/GBM data (see Methods) 
with a spectral index of $0.24\pm0.17$ and a characteristic energy of $67.56\pm11.60$ keV. The final $\chi^{2}$/DoF of 317.81/216 is derived for the fitting. All the error bars represent the 1$\sigma$ statistical uncertainties.
The optical data is brighter than the extrapolation of the gamma-ray spectrum of GRB201223A with an order of about four magnitudes, indicating a distinct radiation mechanism for optical measurement and the high-energy band data. 
}
\label{fig:spectra}
\end{figure}

So far only a small fraction of gamma-ray bursts has been captured in the optical band before the end of the high-energy emission. 
Among these events, very few have been observed during the entire duration of the GRB.
For most GRBs, the prompt optical emission seems to be consistent with or fainter than the extrapolation of high-energy emission to the optical band, consistent with the same synchrotron radiation origin with the $\gamma$-rays\cite{2009MNRAS.398.1936S,2019A&A...628A..59O}. One exception was the naked-eye burst GRB 080319B, which showed a distinct optical emission component with 4 orders of magnitude above the extrapolation of $\gamma$-ray emission\cite{2008Natur.455..183R}. The prompt optical flash detected from GRB 201223A also shows the similar behavior. As shown in Fig.\ref{fig:spectra}, its prompt optical emission is again about 4 orders of magnitude above the extrapolation of $\gamma$-ray emission, even though the flux is fainter by 9 magnitudes. Considering that the redshift of GRB 080319B\cite{2008Natur.455..183R} (z=0.973) is lower than the redshift upper limit of GRB 201223A, the peak prompt optical luminosity of GRB 201223A is likely smaller by more than 2 orders of magnitude than that of GRB 080319B.
Our observation suggests that whatever physical mechanism that operated in the naked-eye GRB (e.g. synchrotron + synchrotron self-Compton\cite{2008Natur.455..183R,2008MNRAS.391L..19K}, emission from residual internal shocks\cite{2008ApJ...674L..65L}, or emission from decayed neutron shells\cite{2009PhRvD..79b1301F}) is not limited to bright events and can apply to moderately bright GRBs as well.

The peak time of the bump indicates the time when the relativistic jet begins to decelerate after interacting with an enough amount of mass in the external medium. Assuming that the burst has a redshift close to the upper limit  (i.e. z$\sim$1.85) and taking the measured $\gamma$-ray spectral parameters, the burst has a typical isotropic $\gamma$-ray energy, $E_{\rm \gamma, iso} \sim 1.8 \times 10^{52}$ ergs (see Methods), with a  high initial Lorentz factor ($\Gamma_{0}\sim 267 \zeta^{1/8} n^{-1/8} $) for its relativistic jet, where $n$ is the external medium density and $\zeta$ is the kinetic-energy-to-$\gamma$-ray energy ratio (see Methods). 
The deceleration radius could be constrained to be  $R_{\rm dec} \sim2.2 \times 10^{17}$ cm.
During the early afterglow phase,
the optical and X-ray emissions share a similar behavior as displayed in Fig.\ref{fig:optical_modeling}, which agrees with the prediction of the standard afterglow theory in the slow cooling regime with $\nu_m < \nu_o < \nu_x < \nu_c$, where $\nu_m$, $\nu_o$, $\nu_x$, $\nu_c$ are the characteristic synchrotron emission frequency with minimum electron Lorentz factor, optical frequency, X-ray frequency,  and synchrotron cooling frequency, respectively. The late X-ray emission is contaminated by a flaring or plateau-like rebrightening behavior which signifies a late-time central engine activity. The common origin of the optical and X-ray early afterglow can be also seen from the SED at 100 - 300 s (Fig.\ref{fig:optical_xray_spec}), which shows that the joint optical-X-ray SED has a spectral index ($\beta_{\rm OX} =2.00\pm0.05$) which is similar to that derived with the X-ray data only ($\beta_{\rm X} = 1.77\pm 0.49$).


It is widely believed that long GRBs originate from the collapse of massive stars\cite{2006ARA&A..44..507W}. 
The leading candidate for the long GRB progenitor is a Wolf-Rayet star \cite{2000ApJ...536..195C}. A major fraction of its mass is expected to be lost in the form of the stellar wind before the collapse. The fact that the ambient medium has a constant-density profile can place an upper limit on the ejected wind mass from the progenitor. For a wind profile with  $\rho = A\, r^{-2}$, where $A=\dot{M}/4\pi V_w = 5 \times 10^{11} A_{*} \, {\rm g\, cm^{-1}}$, the non-detection of a wind medium at early times places a tight upper limit\cite{2004ApJ...606..369C,2009MNRAS.400.1829J} of $A_{*} < 3.4 \times 10^{-2} n^{1/2}$ (see Methods), two orders of magnitude smaller than the typical value expected from a Wolf-Rayet stellar wind with $A_* =1$, the later  corresponds to $\dot{M}=1 \times 10^{-5} {\rm M_{\odot} \, yr^{-1}}$ and $V_w = 1000 \, {\rm km \, s^{-1}}$. Taking the highest wind velocity of $\sim 5000 \ {\rm km \ s^{-1}}$  from known Wolf-Rayet stars in our Galaxy\cite{2000A&A...360..227N}, one can derive a conservative upper limit of the mass loss rate $\dot{M} < 1.7 \times 10^{-6} {\rm M_{\odot} \, yr^{-1}}$, which gives an upper limit of the mass of the progenitor of $\sim$3.8 $M_{\odot}$ with the relation of mass-loss rate and the stellar mass of Wolf-Rayet stars\cite{1989A&A...220..135L} (see Methods). This is in contrast to the naked-eye GRB 080319B, which clearly showed a wind medium and a much larger mass-loss rate and Wolf-Rayet stellar mass. This suggests the mechanism that produces excessively bright prompt optical emission can operate in progenitor systems with a wide-range of mass-loss rate.

The transition from prompt-to-afterglow emission in the optical band was first reported\cite{2006Natur.442..172V}  in GRB 050820A with a long duration of $\sim750$s. The optical observation started 5 seconds after the trigger of its initial precursor, which allowed to detect the superposition between a component that tracks late prompt emission due to continued central engine activity and another smooth component, likely associated with the onset of early afterglow.
For GRB 201223A, the transition from prompt emission to afterglow in the optical wavelength is very smooth without any signature of late central engine activities, and the forward shock emission component before the deceleration of the relativistic outflow was clearly detected immediately after the initial trigger. This offers a new insight into the diversity of the energy release of a GRB jet right after the initial explosion. 
The clear prompt-to-afterglow transition signature in GRB 201223A benefits from the lack of bright reverse shock emission component as observed in some other GRBs\cite{1999Natur.398..400A,2014Sci...343...38V,2017Natur.547..425T}. Physically, this could be due to either the jet is Poynting flux dominated\cite{2005ApJ...628..315Z} or that it is a pure fireball with the reverse shock typical synchrotron frequency $\nu_m$ may below the optical band at shock crossing\cite{2007MNRAS.378.1043J}. For the three possible interpretations to the prompt emission optical excess\cite{2008Natur.455..183R,2008MNRAS.391L..19K,2008ApJ...674L..65L,2009PhRvD..79b1301F}, a matter-dominated ejecta is needed. As a result, a fireball composition gives a unified interpretation to the early optical emission of this event. 

Because GRBs are unpredictable in both time and spatial direction, simultaneous observations of the entire activity of GRBs in both $\gamma$-rays and optical wavelength without any delay are a great challenge even by improving the high slewing-speed of the follow-up telescopes. This limits the progress in studying the physical processes during the prompt emission and early afterglow phases of GRBs. The successful detection of  prompt optical emission right after trigger of this $\sim 29$s-long GRB with GWAC further confirms the feasibility of capturing any bright but short-duration signals from GRBs using large field-of-view instruments. It is foreseeable that prompt optical observation of short-duration originating from neutron star will be made using the similar technique in the near future.





 


\section*{Methods}


\subsection*{Optical observations and data analysis} 

GWAC system is developed for cosmic bright optical transients survey as one of the main ground-based facilities of a China-France satellite mission (SVOM)\cite{2016arXiv161006892W} dedicated to the detection and study of Gamma-ray bursts, aiming to detect various of short-duration astronomical events including the electromagnetic counterparts of gamma-ray bursts by imaging the sky at a cadence of 15 seconds down to $R\sim$16.0 mag, under an automatic observation management\cite{2021PASP..133f5001H}.
A real-time pipeline for short duration transient alert system was developed in GWAC system, named as GWAC transient alert system\cite{2020PASP..132e4502X}.  The method of catalog crossmatching is used to search any short duration transients in real-time pipeline for GWAC data. All  candidates passing through the filters would be followed by two 60cm optical telescopes (F60A and F60B) within two minutes after the alerts, consequently confirmed or rejected automatically by another real-time pipeline developed for F60A/B data\cite{2020PASP..132e4502X}. 
The first four units of GWAC with 16 cameras (D=18 cm) have been built at Xinglong observatory, China. The total field of view is about 2200 square degrees currently. 
The  location of GRB 201223A was monitored during our survey on Dec. 23th, 2012 lasting from  16:48:31.9 to 18:23:41.9 in universal time (UT), which are covering the onset of the burst triggered by Swift satellite (17:58:26 UT). When our system received the alerts, GWAC did not 
do  point adjustment due to its large field of view comparing its high precise localization of the event provided by the instruments of the Swift satellite. The images for this event were not discontinued and 256 white-band  images were obtained  in total. 

We performed the analysis the GWAC images  with a standard aperture photometry at the location of the burst and for
several nearby bright reference stars by using the IRAF\cite{1986SPIE..627..733T}
APPHOT package, including the corrections of bias, dark
and flat-field in a standard manner. 
Since the faint brightness of the transient, a small radius of the aperture for the photometry was selected to be 1$\times$FWHM pixels for each image. The FWHM is the Full width of Half Maximum which is typically used to describe the energy centralization of an image. The FWHM for each frame was  individually measured by the same nearby bright source list yarding a typical value of 1.8 pixels with a variation of 0.1 pixels during our observations. 
After a differential photometry, the finally calibrated brightness of optical counterpart was transformed to $R$ band in Johnson-Cousins system\cite{2005ARA&A..43..293B}.
For the images obtained before the trigger time,  no any signal was detected. We stacked four successive single images to one co-added image to deepen our detection capability. None any credible signals was detected in all the combined images prior to the Swift trigger time. 

\begin{figure}
\centering
\includegraphics[width=1.0\linewidth]{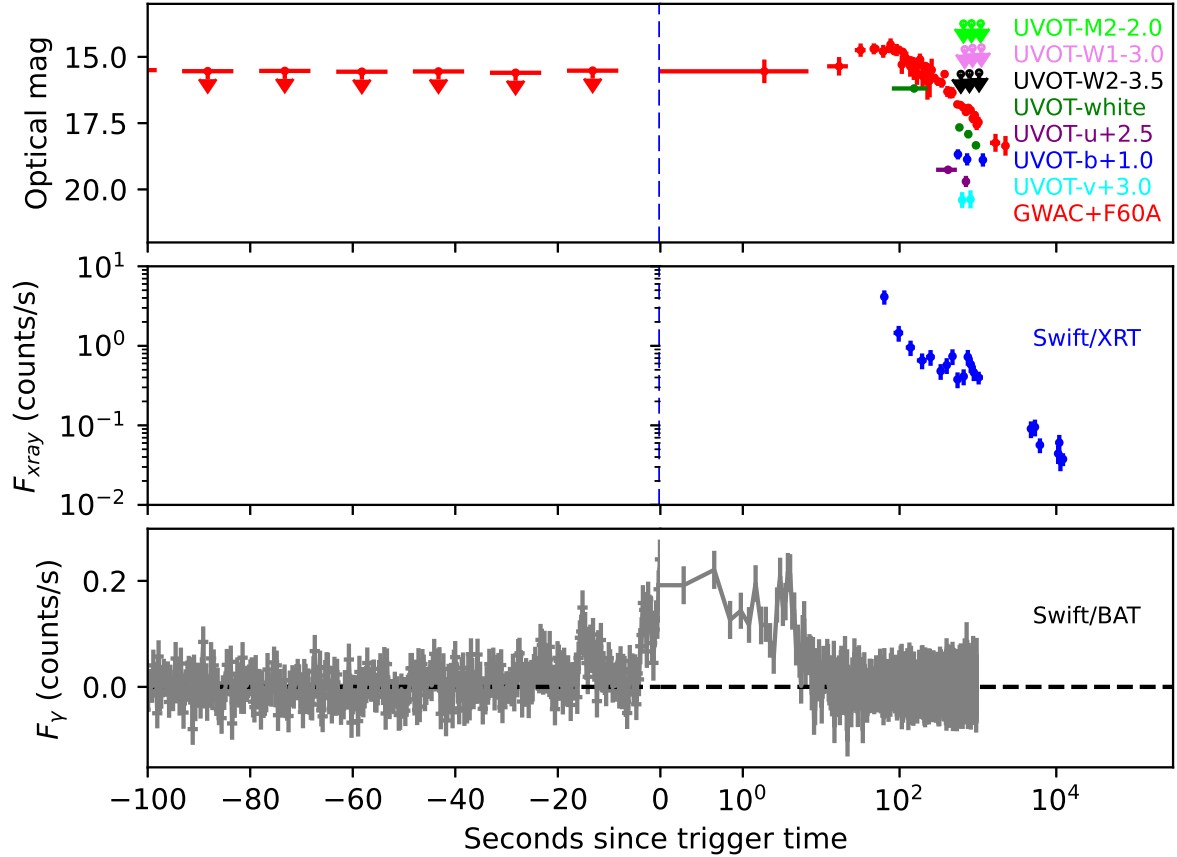}
\caption{\textbf{A comparison of the prompt $\gamma$-ray, X-ray and optical light curves GRB 201223A measured by Swift/BAT, Swift/XRT, Swift/UVOT, GWAC and F60A from the time before the event to $\sim10^{4}$ seconds after the Swift trigger time.}  The BAT emission from 15-150 kev have been integrated over 0.256 seconds time interval. The horizontal lines for each individual data denote the observing intervals and the vertical lines represent the 1-sigma error bars. The lower arrows represent the upper limit for these measurements. The first optical emission was captured  coincidentally during the main gamma-ray pulse. The  figure is separated into two parts: The X-axis in the left panels before the trigger time are displayed in linear space, while the temporal axis in the right panels after the trigger time are shown in the logarithmic space.   The measured optical light curves have four segments. No optical emission was detected from the source before the trigger time. After the peak time of the prompt $\gamma$-ray emission, there is a short-living plateau or a shallow increasing optical emission in the first $\sim$17 seconds. After that, a rising feature was detected with a temporal pow-law index of s$\sim 0.66\pm0.30$ to the peak time of t$_p \sim 52\pm30$ seconds, followed by a late temporal pow-law decaying with an index of $\sim 1.07\pm0.15$. The X-ray emission shows a similar behavior with optical light curve. 
More information of the statistics is presented in Fig.\ref{fig:optical_modeling}. }
\label{fig:lc}
\end{figure}

The narrow-field optical telescope F60A  has a rapid follow-up system for gamma-ray bursts triggered by Swift BAT instrument. 
This telescope is located near the GWAC facilities and jointly operated by  Guangxi university and National Astronomical Observatories, CAS, China. 
The Diameter of the telescope is 60cm, and its field of view is 19 arc-minutes and the pixel scale is 0.56 arc-seconds 
when 2K$\times$2K astronomical CCD detector is equipped. 
When GRB 220213A was triggered, F60A pointed to the location of the alert rapidly 
and started to image the sky at 45 seconds after the onset of the event. 
A series of $R$ and $I$ images were obtained with a pre-defined strategy 
depending on how much of the delay for the first measurement. 
All the images are processed with a standard manner with a correction of the bias, 
dark and flat-field. All the images are aligned with a reference of the bright source list in the images. 
A aperture photometry was utilized with the same method for GWAC images. 
Since the decaying nature of the optical afterglow, the image obtained 
at later time were combined to increase the signal to noise ratio. 
In the Fig.\ref{fig:lc}, the $I$ magnitudes was shifted to $R$ band by adding 0.2 mag to normalize the optical light curves. 

A smoothed broken power-law model, i.e.
\begin{equation}
    f=f_0 \times \left[ \left(\frac{t}{t_b}\right)^{w a_1} + \left(\frac{t}{t_b}\right)^{w a_2} \right]^{\frac{1}{w} } 
    \label{equ:bpl}
\end{equation}
was utilized to model the optical light curve as a result shown in Fig.\ref{fig:lc}, where $t$ is the time (in units of seconds) since the Swift/BAT trigger, $f_0$ is the normalization constant, $\alpha_1$ and $\alpha_2$ are the temporal decay indices, respectively, $t_b$ is the broken time, and $w$ is the smooth parameter.

\subsection*{Swift and Fermi/GBM data analysis}

We downloaded the Swift/BAT data for GRB 201223A from the Swift archive website.
The $batbinevt$ was used to extract the total light curve.  
The time-integrated spectrum near the Swift high-energy peak from -2.31 to 7.69 seconds after the Swift trigger time was  derived.  
We also downloaded GRB 201223A archival data from the Fermi/GBM website.  
The standard analysis was carried out with the {\em Fermi} data analysis tools and the HEAsoft packages. 
GRB 201223A was detected by several detectors. We chose the data from the NaI detectors of $n7$, $n8$, and the BGO detector $b1$ based on the signal-to-noise ratios of each detector. 
The lightcurve (The Supplementary Fig.\ref{fig:gbm_lc}) in each detector relative to the {\em Swift}/BAT trigger time was extracted with $gtbin$. The background was selected in the time range of [-60s:-20s] and [100s:140s] relative to the {\em Swift} trigger time. A joint analysis was performed via Xspec12\cite{1996ASPC..101...17A} for the GWAC data, the {\em Swift}/BAT data and the {\em Fermi}/GBM data with three models: the single power-law model 
\begin{equation}
    A(E) = K \times E^{-\alpha },
    \label{equ:pl}
\end{equation}
where $E$ is the energy in keV. $\alpha$ is the power law photon index and $K$ is the normalization in $photons/keV/cm^{2}/s$ at 1 keV; the cutoff power-law model 
\begin{equation}
    A(E) = K \times E^{-\alpha } e^{-E/\beta}, 
    \label{equ:cpl}
\end{equation}
where $E$ is the energy in keV, $\alpha$ is the power law photon index, $\beta$ is the e-folding energy of the exponential rolloff (in keV), and $K$ is the normalization in $photons/keV/cm^{2}/s$ at 1 keV; and the Band function\cite{1993ApJ...413..281B} model
\begin{equation}
    A(E)=\left\{\begin{array}{ll}
K(E / 100 .)^{\alpha_{1}} \exp \left(-E / E_{c}\right) & \text { if } E<E_{c}\left(\alpha_{1}-\alpha_{2}\right) \\
K\left[\left(\alpha_{1}-\alpha_{2}\right) E_{c} / 100\right]^{\left(\alpha_{1}-\alpha_{2}\right)}(E / 100)^{\alpha_{2}} \exp \left(-\left(\alpha_{1}-\alpha_{2}\right)\right) & \text { if } E>E_{c}\left(\alpha_{1}-\alpha_{2}\right)
\end{array}\right. ~,
\label{equ:BandFunction}
\end{equation}
where $E$ is the energy in units of keV, $\alpha_1$ and $\beta_1$ are the low-energy and high-energy spectral indices, $E_c$ is the characteristic energy in keV, and $K$ is the normalization constant. The fitting results (Supplementary data Table.1) suggest that the cutoff power-law model is the best model, and the result from this model is displayed in Fig.\ref{fig:spectra}.

\begin{figure}
\centering
\includegraphics[width=1.0\linewidth]{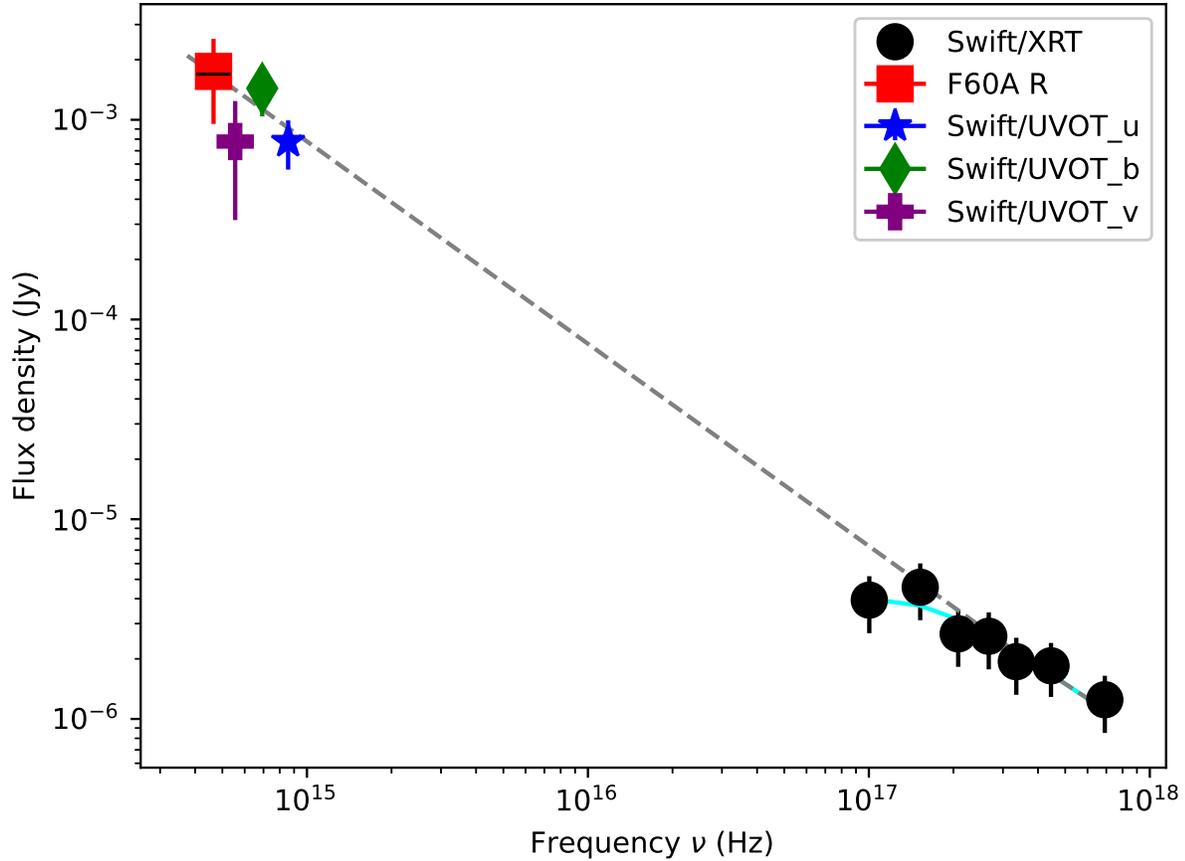}
\caption{\textbf{The spectral energy distribution between X-ray and optical wavelengths during the time window from 100 sec to 300 sec after the Swift/BAT trigger time.} The X-ray spectrum in black is extracted from Swift/XRT data. The red data is measured from GWAC data.
The joint fitting with GWAC $R$-band data and X-ray data gives a photon index of $\beta_{OX}=2.00\pm0.05$ with a $\chi^{2}/DoF$=3.47/6 after taking  a photo-electric absorption (wabs model) into account using the X-ray Spectral Fitting Package (Xspec12)\cite{1996ASPC..101...17A}, which agrees with that derived by fitting the X-ray data only ($\beta_X = 1.77\pm0.49$ with a $\chi^{2}/DoF$=4.05/5 ).  The fitting result with the total model (wabs + powerlaw) is shown in cyan and the only power-law component is displayed in gray dash line. 
The UVOT data were also displayed which were not included during the fitting (see Methods and Supplementary data Figure 1). 
All the crosses for each data represent the 1$\sigma$ errors. 
}
\label{fig:optical_xray_spec}
\end{figure}

\begin{figure}
\centering
\includegraphics[width=1.0\linewidth]{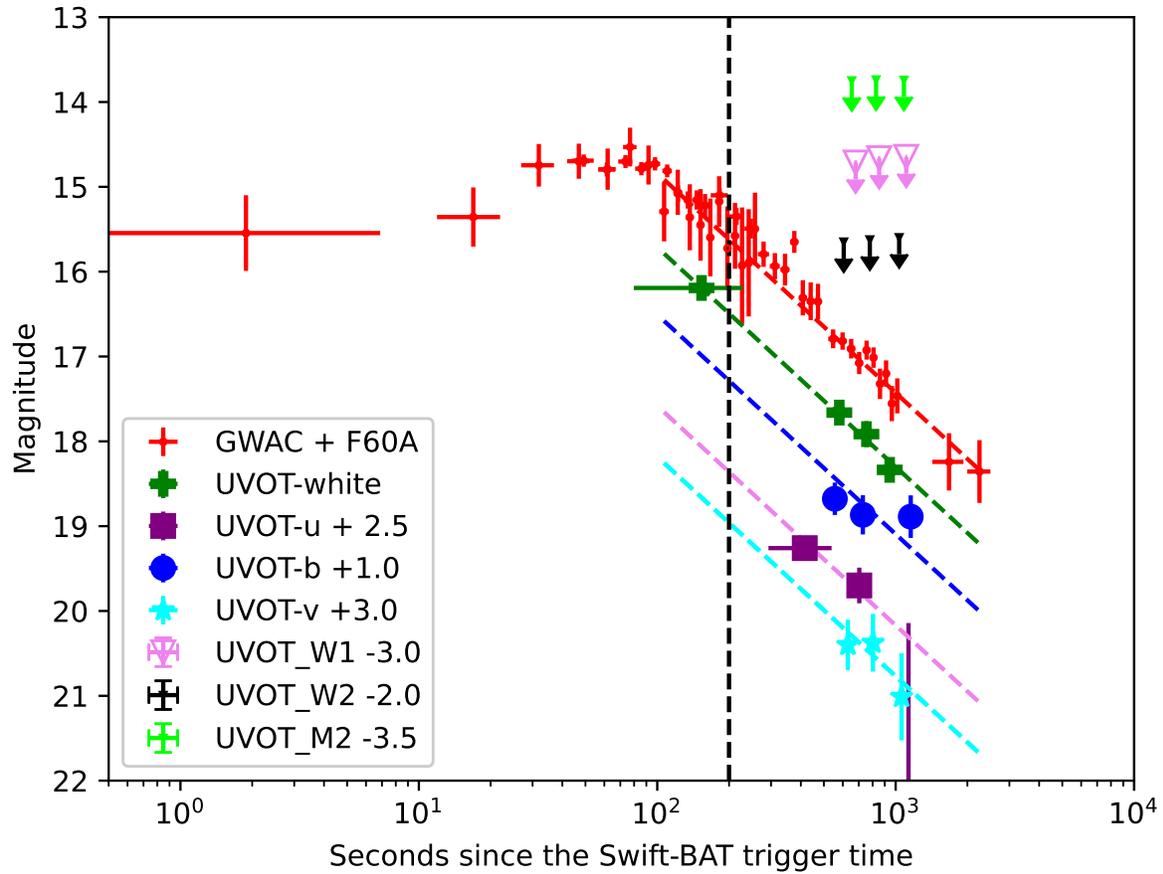}
\renewcommand{\figurename}{Supplementary Figure} 
\renewcommand{\thefigure}{1}
\renewcommand*{\thefigure}{S\arabic{figure}}
\caption{
\textbf{Optical light curve measured by GWAC, F60A and Swift/UVOT, and its modeling.} A single power-law model is used to fit the data obtained with GWAC and F60A after 100 seconds since the trigger time, resulting a decay index of $\alpha=1.03\pm0.03$ with a $\chi^{2}$ of 16.79 and a degree of freedom (DoF) of 26.
The UVOT data is modeled by fixing the same decay index as that of F60A,  with an assumption that a achromatic decay among these wavelength during the forward shock phase. 
The fitting results the $\chi^{2}/DoF$ of 1.57/3, 0.52/2, 1.69/1, 4.1/2 for UVOT-white, UVOT-v, UVOT-u and UVOT-b, respectively. 
All the Swift/UVOT data are shifted for clarity. The black dashed line shows 200 seconds after the Swift trigger time.
The 1$\sigma$ errors are denoted by the crosses for each measurement.
}
\label{fig:gwac_uvot_lc_fitting}
\end{figure}

The {\em Swift}/XRT data are downloaded from the {\em Swift} data archives. The lightcurve (Fig.\ref{fig:optical_modeling}) and spectrum (Fig.\ref{fig:optical_xray_spec}) at the time epoch between 100 seconds and 300 seconds post-trigger was extracted via the HEAsoft packages and the {\em Swift} data analysis tools. The {\em Swift}/UVOT data for GRB 201223A was downloaded from the {\em Swift} archives website. The standard data products were obtained. 
The photometries in each filter were derived to build the multi-wavelength lightcurves (Fig.\ref{fig:lc} and the Supplementary Fig.\ref{fig:gwac_uvot_lc_fitting}).
The UVOT data was modeled with an assumption of an
achromatic decay in optical wavelengths during the forward shock phase (the Supplementary  Fig.\ref{fig:gwac_uvot_lc_fitting}), which resulted in the predicted brightness in {\em Swift}/UVOT-u, {\em Swift}/UVOT-b and {\em Swift}/UVOT-v filters at 200 seconds after the {\em Swift} trigger time (Fig.\ref{fig:optical_xray_spec}). During the procedure, UVOT-white-band data was excluded for its broad wavelength coverage, and the data from UVOT-W1, UVOT-W2 and UVOT-M2 were not considered for their non-detection (Fig.\ref{fig:lc} and the Supplementary Fig.\ref{fig:gwac_uvot_lc_fitting}).

\subsection*{Isotropic energy} The parameters of the time-integrated spectrum were adopted from the Fermi/GBM data\cite{2020GCN.29161....1W}. With the same method\cite{2001AJ....121.2879B}, the isotropic energy could be estimated by integration of the GBM spectrum from 1 keV to 10000 keV in the burst frame, under an upper limit of the redshift z $\sim$ 1.85 based on the detection of the Swift/UVOT with a blue filter\cite{2020GCN.29158....1G}. The isotropic $\gamma$-ray energy is estimate to be $E_{\gamma,kiso} \simeq 1.8 \times 10^{52}$ ergs.

\subsection*{Initial Lorentz factor and the deceleration radius}  One can calculate the initial Lorentz factor with the peak time of the onset of the afterglow\cite{2018pgrb.book.....Z},
$\Gamma_{0} \simeq 0.9^{3/8} \left(\frac{3E_k (1+z)^{3}}{2 \pi \hat{\gamma} n m_{p} c^{5} t_{dec}^{3}}\right)^{1 / 8}  \simeq 170 t_{dec,2}^{-3/8} (\frac{1+z}{2} )^{3/8} E_{k,52}^{1/8} n^{-1/8}$ , where $E_k$ is the total isotropic kinetic energy in the fireball, and $t_{dec}$ is the onset of the afterglow when the relativistic jet starts to decrease, which should be near the peak time of the bump. Taking $E_{\gamma,iso}=1.8\times10^{52}$ erg, and $E_k=\zeta E_{\gamma,iso}=\frac{1-\eta_{\gamma}}{\eta_{\gamma}}E_{\gamma,iso}$, where $\eta_{\gamma}=E_{\gamma, \text { iso }}/(E_{\gamma, \text { iso }}+E_{K, \text { iso}})$ is the GRB radiative efficiency, one gets $\Gamma_{0} \simeq 233 \zeta^{1/8} n^{-1 / 8} (\frac{1+z}{2} )^{3/8} \simeq 267 \zeta^{1/8} n^{-1 /8}$ when $z=1.85$ is adopted, where $n$ is the external medium density. 
Consequently, the deceleration radius $R_{\rm dec} \sim 2 c\Gamma_{0}^{2} t_{dec} (1+z)$ would be  $\sim 2.2\times10^{17}$ cm for typical parameters ($\eta_{\gamma}=0.5$, $\zeta=1$, $n=1$ and $z=1.85$).

\subsection*{Constraint on the parameter of stellar wind density}
Assuming there is a stellar wind due to the massive stars, the wind-to-ISM transition time $T_r$ at transition radius $R_t$ can be estimated\cite{2004ApJ...606..369C}$^{,}$ \cite{2009MNRAS.400.1829J} with the relation
$T_{r}=1.5 \mathrm{~h}\left(\frac{1+z}{2}\right) E_{\mathrm{k}, 53}^{-1} A_{*,-1}^{2} n^{-1} \sim 50 \mathrm{~s}$.
Thus, we have $A_{*,-1}^{2} n^{-1} \sim 6.5\times10^{-3}E_{\mathrm{k}, 53} $.
$A_{*}$ is a parameter of  stellar wind density\cite{1998A&A...333L..87D}$^{,}$\cite{2000ApJ...536..195C} scaled with the typical values in terms of the wind from a Wolf-Rayet star.
The GRB radiative efficiency defined as $\eta_{\gamma}=E_{\gamma, \text { iso }}/(E_{\gamma, \text { iso }}+E_{K, \text { iso}})$ has a typical value $\eta_{\gamma}$\cite{2007ApJ...655..989Z,2016MNRAS.461...51B,2015ApJS..219....9W,2020ApJ...900..176L} from 3\% to $\sim90$\%.
Assuming $\eta_{\gamma} \sim 50\%$ for GRB 201223A,  E$_\mathrm{k} \sim1.8\times10^{52}$ ergs would be obtained. Consequently, an upper limit could be inferred as 
$A_{*} \sim 3.4 \times 10^{-2} n^{1/2}$.

With the definition of the stellar wind density 
$A=\dot{M}/4\pi V_w = 5 \times 10^{11} A_{*} \, {\rm g\, cm^{-1}}$, one note that for a certain $A_{*}$, more higher a velocity $V_w$, more larger the mass loss rate $\dot{M}$. Based on the study\cite{2000A&A...360..227N}, the velocities of the stellar wind for the Wolf-Rayet stars in our Galaxy are distributed in the range between 700 km s$^{-1}$ and 5000 km $s^{-1}$. Taking the largest velocity of 5000 km s$^{-1}$, the upper limit of the mass of the progenitor could be estimated to be 3.8$M_{\odot}$
based on the relation\cite{1989A&A...220..135L,1999ApJ...520..641S} between the mass loss rate and the mass of a Wolf-Rayet star
$\dot{M} \sim 6\times 10^{-8} \times (\frac{M_{WR}}{M_{\odot }})^{2.5}  M_{\odot} yr^{-1}$.

\section*{Data Availability}
Data generated or analysed during this study are included in this Article (and its Supplementary
Information). Source data are provided with this paper. 

\section*{Code Availability}
The analysis codes used to generate the data presented in this study are available from the
corresponding authors upon reasonable request.

\section*{Acknowledgements}
This study is supported from 
the National Natural Science Foundation of China (Grant No. 11973055, U1938201, 12133003, U1831207, U1931133) and partially supported by the Strategic Pioneer
Program on Space Science, Chinese Academy of Sciences, grant
Nos. XDA15052600 and XDA15016500. 
JW is supported by the National Natural Science Foundation of
China (Grants No. 12173009), and the Natural Science Foundation of Guangxi (2020GXNSFDA238018).
XYW is supported by the National Natural Science Foundation of China under grants 12121003. 
YGY is supported by the National Natural Science Foundation of China under grants 11873003. 
This work made use of data supplied by the UK Swift Science Data Centre at the University of Leicester.

\section*{Author Contributions Statement}


LX led the project and paper writing.  HL, LX, JW, CW, HC, YQ reduced and analyzed the optical data. LX, DT, LZ and XY analyzed the high-energy data. XH, XL, LH performed GWAC and F60A observations. BZ, LX, JD, HG and JR presented the interpretation to the data and BZ contributed to paper writing. EL, XW, ZD, XW, and YY partially funded the facilities. JYW is the PI for the GWAC GRB project.  All authors reviewed the manuscript.

\section*{Competing Interests Statement}

The authors declare that they have no competing financial interests. 
Correspondence and requests for materials should be addressed to Liping Xin (xlp@nao.cas.cn), Jianyan Wei (wjy@nao.cas.cn) and Bing Zhang(bing.zhang@unlv.edu). 

\section*{Tables}

\section*{Figure Legends/Captions (for main text figures)}

\renewcommand{\figurename}{Extended Data Figure} 
\renewcommand{\tablename}{Supplementary Table} 
\begin{figure}
\centering
\includegraphics[width=1.0\linewidth]{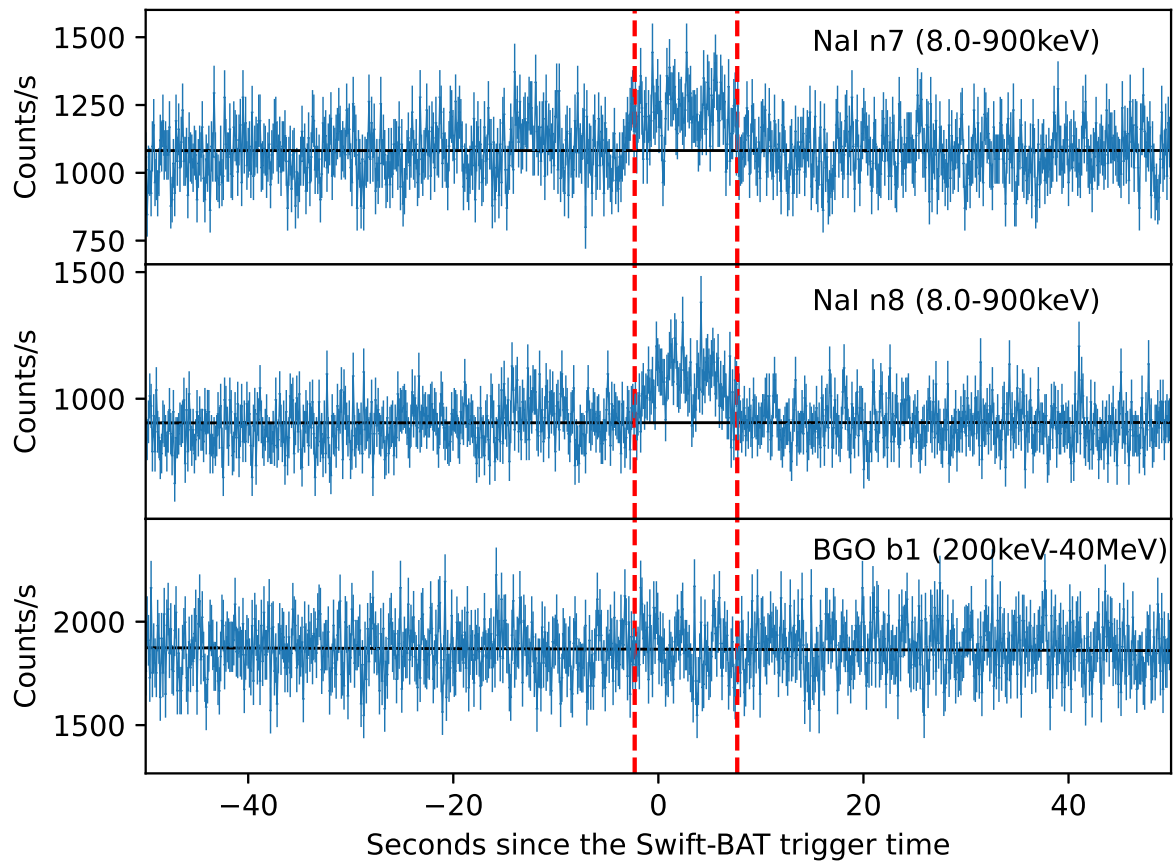}
\renewcommand{\figurename}{Supplementary Figure} 
\renewcommand{\thefigure}{2}
\caption{
\textbf{High-energy light curves of GRB 201223A derived from Fermi/GBM in three detectors.} The x-axis is the seconds relative to the Swift/BAT trigger time.
The y-axis is the counts per second. The black line in each panel shows the background emission level in each detector. The time window [-2.31s:7.69s] in two dash lines is the effective exposure time of GWAC for the prompt optical emission. 
}
\label{fig:gbm_lc}
\end{figure}

\clearpage

\begin{table}
\begin{center}
\begin{tabular}{cccccc} 
\hline 
Model  &   $\mathrm{\alpha}$($\alpha_1$)  & $\mathrm{\beta}$($\alpha_2$) & $\mathrm{K}$  & $E_{c}(keV)$ & $\chi^{2}/DoF$\\
\hline 
Power law &  1.32$\pm$0.01   &   &  1.15$\pm$0.06  &  & 554.48/217 \\
Cutoff power law  &  0.24$\pm$0.17  &    &  0.06$\pm$0.03 &  67.56$\pm$11.60 & 317.81/216\\
The Band function &  -0.01$\pm$0.24 & -2.3(fixed) & 0.03$\pm$0.01 & 50.46$\pm$12.37 & 319.84/216 \\ 
\hline
\label{Tab:model_fitting}
\end{tabular}
\end{center}
\renewcommand{\tablename}{Supplementary Table} 
\caption{\textbf{Modeling results for the spectral energy distribution shown in Fig.\ref{fig:spectra} during the prompt emission.} $\mathrm{\alpha}$($\alpha_1$) and $\mathrm{\beta}$($\alpha_2$) are the low-energy and high-energy spectral indices, $E_{c}$ is the characteristic
energy in keV, and K is the normalization constant (see Methods). $DoF$ is the degree of freedom during the fitting.
The errors for each values represents the 1$\sigma$ errors.}
\end{table}


\end{document}